\documentclass[review]{elsarticle}
\usepackage{amsmath}

\journal{Journal of \LaTeX\ Templates}

\begin{document}

\begin{frontmatter}

\title{Unconventional photon blockade based on double second-order nonlinear coupling system}

\author{Hongyu. Lin$^{1,2}$, Xiaoqian. Wang$^{1}$, Zhuo. Yu$^{1}$ and Zhihai. Yao$^{1*}$ }
\address{$^{1}$Department of Physics, Changchun University of Science and Technology, Changchun 130022, China}
\address{$^{2}$ College of Physics and electronic information, Baicheng Normal University, Baicheng 137000, China}

\cortext[Zhihai. Yao]{Corresponding author}
\ead{yaozh@cust.edu.cn}
\begin{abstract}
In the recent publications [Phys. Rev. A 92,
023838 (2015)], the unconventional photon blockade are studied in a
two-mode-second-order-nonlinear system with nonlinear coupling
between the low frequency and high frequency modes. In this paper, we study the
unconventional photon blockade in a three-mode system with weakly coupled nonlinear cavities via $\chi^{(2)}$ nonlinearity. By
solving the master equation in the steady-state limit and
calculating the zero-delay time second-order correlation function, we obtain the conditions of strong photon antibunching in the low frequency mode. The numerical result are compared with the analytical results, the results show that they are in complete agreement.
By the analysis of numerical solutions, we find that this scheme is not sensitive to the change of decay rates and the reservoir temperature, and the three-mode drives make the system have more adjustable parameters, both of which increases the possibility of experimental implementation.
\end{abstract}


\end{frontmatter}

\section{Introduction}
With the rapid development of quantum communication~\cite{04}, quantum metrology~\cite{05}, quantum information technology~\cite{06} and other fields, single photon sources~\cite{02,03} has become a hot topic in modern scientific research due to its huge application potential in these fields. The realization of single photon source mainly depends on the systems driven under a classical light field
can producing sub-Poissonian light. Based on the above physical mechanism a method called photon blockade (PB)~\cite{10,11} is gaining popularity. Photon blockade is a strong antibundling phenomenon, when a nonlinear resonator produces a photon, it will block the generation of the second photon. It has been proved experimentally that PB can be realized in cavity-QED~\cite{12} or circuit-QED~\cite{13} systems. And according to the prediction, PB has potential application value in many fields such as nonlinear optical systems~\cite{14,15,16} and optomechanical devices~\cite{17,18}. There are two main mechanisms to realize PB, one is the large energy levels splitting due to the nonlinearity of the system, which is called conventional photon blockade (CPB)~\cite{19,z}, and the other is a strong photons anticlustering caused by quantum interference, which is called unconventional photon blockade (UPB)~\cite{ab}.
Now the CPB has been implemented on many systems, such as cavity quantum electro dynamics~\cite{20}, quantum optomechanical systems~\cite{22,24,25} and semiconductor micro cavities with second-order nonlinearity~\cite{26,27,e}. In addition, CPB has potential applications in single-photon transistors~\cite{28}, interferometers~\cite{29}, and quantum optical rectifiers~\cite{30,31}.

The UPB was recently discovered by Liew and Savona~\cite{ab} in two semiconductor microcavities~\cite{33,34,c} with weak nonlinear coupling, then the UPB implemented in many similar weakly nonlinear coupled systems~\cite{ab,b,d,f}. With the development of this field, and the mechanism is universal, many different nonlinear systems are proposed to realize UPB, such as including bimodal optical cavity with a quantum dot~\cite{37,38}, coupled polaritonic systems~\cite{39}, coupled optomechanical systems~\cite{40}, or coupled single-mode cavities with second-order or third-order nonlinearity~\cite{42,43,44}. In addition to single-photon control, UPB can also be used as a tool to reveal non-classical features, including the booming development of semiconductor microcavities~\cite{45}, the search for quantum correlations~\cite{46}. In addition, the observed non-classical optical statistics of exciton-polaritron field also depend on UPB~\cite{47}. Recently, in Ref.~\cite{01}, the unconventional photon blockades are studied in a two-mode-second-order-nonlinear system with nonlinear coupling between the low frequency and high frequency modes, and the conditions of strong photon antibunching is obtained in the low frequency mode.

In this paper, we investigate the unconventional photon blockade in a three-mode system contains double second order nonlinear coupling.
The optimal condition for strong antibunching is found in the low frequency mode by analyticcal culations and discussions of the optimal condition are presented. By comparing the numerical and analytical solutions, we find that they are in good agreement. By the analysis of numerical solutions, we find that this scheme is not sensitive to the change of decay rates and the reservoir temperature, and the three-mode drives make the system have more adjustable parameters, and in the experiments, the relatively large blocking windows are required~\cite{i,j,n}, the form of a tri-mode drive system can provide more tunable parameters to form a larger blocking window, both of which increases the possibility of experimental implementation.

The remainder of this paper is organized as follows: In Sec.~2, we introduce the physical model of the double second-order nonlinear coupling system. In Sec.~3, by analytical calculation, we derive an expression for the optimized antibunching, and we give a numerical result and compare it with the analytical one. Conclusions are given in Sec.~4.

\section{Physical model}
\begin{figure}[h]%"[]"中为位置参数，四个参数tbph 依次是置顶、置底 、浮动、当前位置，，选用的参数优先顺序为h-t-b-p
\centering
\includegraphics[scale=0.50]{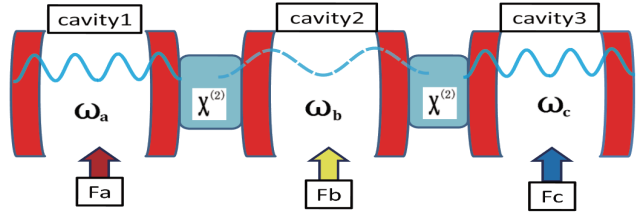}%"scale"后的数字为图形的宽度，也可用"width=1.0\columnwidth"定义
\caption{The scheme of the system consists of two low frequency cavities 1 and 3 with frequencies are $\omega_a$ and $\omega_c$, and one low frequency cavity 2 with frequency is $\omega_b$. The coupling between cavities is by $\chi^{(2)}$ nonlinear material. $F_a$, $F_b$ and $F_c$ are the driving strength for mode $a$, $b$ and $c$, respectively.}
\label{fig1}
\end{figure}

In thie paper we study the unconventional photon blockade in double second order nonlinear coupling system, which consists of two high frequency cavities $a$ and $c$ with frequencies are $\omega_a$ and $\omega_c$, and one low frequency cavity $b$ with frequency is $\omega_b$. The three cavities are coupled by two-order-nonlinear $\chi^{(2)}$ materials that mediates the conversion of the single-photon in cavity 1 or cavity 3 into two-photon in cavity 2, the system model diagram is shown in Fig.~\ref{fig1}.
External weak drive is the key to realize PB, and in this system we chose to three-mode drives mode. The Hamiltonian of the system can be written as Ref.~\cite{01,50}
\begin{eqnarray}
\hat{H}&=&\omega_a\hat{a}^{\dag}\hat{a}\
+\omega_b\hat{b}^{\dag}\hat{b}+\omega_c\hat{c}^{\dag}\hat{c}\
+J(\hat{a}^\dag\hat{b}^2+\hat{b}^{\dag2}\hat{a})
+g(\hat{c}^{\dag}\hat{b}^2+\hat{b}^{\dag2}\hat{c}) \nonumber\\
&&+F_a(\hat{a}^{\dag}e^{-i\omega_Lt}+\hat{a}e^{i\omega_Lt})
+F_b(\hat{b}^{\dag}e^{-i\omega_Lt}+\hat{b}e^{i\omega_Lt})
+F_c(\hat{c}^{\dag}e^{-i\omega_Lt}+\hat{c}e^{i\omega_Lt}),
\label{01}
\end{eqnarray}
Here the $\hat{a}(\hat{a}^{\dag})$, $\hat{b}(\hat{b}^{\dag})$ and $\hat{c}(\hat{c}^{\dag})$ denotes the annihilation (creation) operator of the three cavities, the $F_a$, $F_b$ and $F_c$ express the driving strength for the three cavities, $\omega_L$ is the driving frequency, the $g$ and $J$ express the second order nonlinear strengths, which can be derived from the $\chi^{(2)}$ nonlinearity as~\cite{26}
\begin{eqnarray}
g=D\varepsilon_0(\frac{\omega_a}{2\varepsilon_0})\sqrt{\frac{\omega_b}{2\varepsilon_0}}\int d\textbf{r}\frac{\chi^{(2)}(\textbf{r})}{[\varepsilon(\textbf{r})]^{3/2}}\alpha_a^2(\textbf{r})\alpha_b(\textbf{r}).
\label{02}
\end{eqnarray}
\begin{eqnarray}
J=D\varepsilon_0(\frac{\omega_c}{2\varepsilon_0})\sqrt{\frac{\omega_b}{2\varepsilon_0}}\int d\textbf{r}\frac{\chi^{(2)}(\textbf{r})}{[\varepsilon(\textbf{r})]^{3/2}}\alpha_c^2(\textbf{r})\alpha_b(\textbf{r}).
\label{03}
\end{eqnarray}
The $\varepsilon_0$ and $\varepsilon_r$ express the vacuum permittivity and relative permittivity, respectively. The $\alpha_a(\textbf{r})$, $\alpha_b(\textbf{r})$ and $\alpha_c(\textbf{r})$ are the wave functions for mode $a$, mode $b$ and mode $c$.
According to the rotation
frame work operator $\hat{U}(t)=e^{i t(\omega_{l}\hat{a}^{\dag}\hat{a}+\omega_{l}\hat{c}^{\dag}\hat{c}+\omega_{l}\hat{b}^{\dag}\hat{b})}$,
We can get an effective Hamiltonian
$\hat{H}_{eff}=\hat{U}\hat{H}\hat{U}^{\dag}-i \hat{U}dU^{\dag}/dt$
as
\begin{eqnarray}
\hat{H}_{eff}&=&\Delta_a\hat{a}^{\dag}\hat{a}\
+\Delta_b\hat{b}^{\dag}\hat{b}+\Delta_c\hat{c}^{\dag}\hat{c}+J(\hat{a}^\dag\hat{b}^2+\hat{b}^{\dag2}\hat{a}) \nonumber\\
&&+g(\hat{c}^{\dag}\hat{b}^2+\hat{b}^{\dag2}\hat{c})
+F_a(\hat{a}^{\dag2}+\hat{a})+F_b(\hat{b}^{\dag}+\hat{b})+F_c(\hat{c}^{\dag}+\hat{c}),
\label{04}
\end{eqnarray}
Where $\Delta_a=\omega_a-\omega_L$, $\Delta_b=\omega_b-2\omega_L$ and $\Delta_c=\omega_c-\omega_L$ represent the detunings of the cavity $a$, $b$ and $c$, respectively. Experimental implementations of similar models have been discussed in Refs.~\cite{27,53}.
The dynamics of the density matrix $\rho$ of the system is governed by
\begin{eqnarray}
\frac{\partial\hat{\rho}}{\partial t}=-i[\hat{H},\hat{\rho}]+\kappa_a\ell(\hat{a})\hat{\rho}+\kappa_b\ell(\hat{b})\hat{\rho}+\kappa_c\ell(\hat{c})\hat{\rho},
\label{05}
\end{eqnarray}
Here the $\kappa_a$, $\kappa_b$ and $\kappa_c$ represent the damping constants of
cavity $a$, $b$, and $c$, respectively. The super operator
is defined by $\ell(\hat{o})\hat{\rho}=\hat{o}\hat{\rho}\hat{o}^{\dag}-\frac{1}{2}\hat{o}^{\dag}\hat{o}\hat{\rho}-\frac{1}{2}\hat{\rho}\hat{o}^{\dag}\hat{o}$.
For convenient to calculate we set $\kappa_a=\kappa_b=\kappa_c=\kappa$.
In this paper, we will analyze the UPB in the low-frequency mode $b$, and we use the steady-state second-order correlation function to describe the statistical properties of photons, which can be calculated by solving the master
equations numerically as
\begin{eqnarray}
g^{(2)}(0)=\frac{\langle
\hat{b}^{\dag}\hat{b}^{\dag}\hat{b}~\hat{b}\rangle}
{\langle\hat{b}^{\dag}\hat{b}\rangle^2}, \label{06}
\end{eqnarray}
The $g^{(2)}(0)\ll 1$ indicates that the UPB occurs in mode $b$.
\section{Analytical and numerical results}
Next, we will analyze the system analytically, according to the system Hamiltonian, the wave function can be written as
\begin{eqnarray}
|\psi\rangle &=& C_{000}|000\rangle+C_{100}|100\rangle+C_{010}|010\rangle+C_{001}|001\rangle\nonumber\\
&&+C_{020}|020\rangle.
\label{07}
\end{eqnarray}
Here $|mnp\rangle$ denotes the Fock-state basis of the system with the number $m$, $n$ and $p$ denoting the photon number in mode $a$, $b$, and $c$, respectively.
For the system, which containing two high frequency mode photons, this is a good approximation. Consider environmental influence we can  treat the system by the non-Hermitian Hamiltonian
\begin{eqnarray}
\widetilde{H}=\hat{H}_{eff}-i\frac{\kappa}{2}\hat{a}^\dag\hat{a}
-i\frac{\kappa}{2}\hat{b}^\dag\hat{b}-i\frac{\kappa}{2}\hat{c}^\dag\hat{c}
\label{08}
\end{eqnarray}
The $\hat{H}$ is given in Eq.~(\ref{01}).
Substituting the wave function Eq.~(\ref{07})
and non-Hermitian Hamiltonian into the Schr\"odingers equation
$i\partial_t|\psi\rangle=\widetilde{H}|\psi\rangle$,
we can obtain the coupled equations for the coefficients
\begin{eqnarray}
&&i\dot{C}_{000}=F_b C_{010}+F_a C_{100}+F_c C_{001}\nonumber\\
&&i\dot{C}_{100}=F_a C_{000}+(\Delta_a-\frac{i\kappa}{2})C_{100}+\sqrt{2}J C_{020},\nonumber\\
&&i\dot{C}_{010}=F_b C_{000}+(\Delta_b-\frac{i\kappa}{2})C_{010}+\sqrt{2}F_b C_{020},\nonumber\\
&&i\dot{C}_{001}=F_c C_{000}+(\Delta_c-\frac{i\kappa}{2})C_{001}+\sqrt{2}gC_{020},\nonumber\\
&&i\dot{C}_{020}=\sqrt{2}F_b C_{010}+\sqrt{2}J C_{100}+\sqrt{2}g C_{001}+2(\Delta_a-\frac{i\kappa}{2})C_{020}
\label{09}
\end{eqnarray}
Therefore, the steady-state coefficient equation can be expressed as
\begin{eqnarray}
&&F_b C_{010}+F_a C_{100}+F_c C_{001}=0,\nonumber\\
&&F_a C_{000}+(\Delta_a-\frac{i\kappa}{2})C_{100}+\sqrt{2}J C_{020}=0,\nonumber\\
&&F_b C_{000}+(\Delta_b-\frac{i\kappa}{2})C_{010}+\sqrt{2}F_b C_{020}=0,\nonumber\\
&&F_c C_{000}+(\Delta_c-\frac{i\kappa}{2})C_{001}+\sqrt{2}gC_{020}=0,\nonumber\\
&&\sqrt{2}F_b C_{010}+\sqrt{2}J C_{100}+\sqrt{2}g C_{001}+2(\Delta_a-\frac{i\kappa}{2})C_{020}=0,
\label{10}
\end{eqnarray}
Because to implement UPB the weak driving strength must be satisfied, namely the $F_a, F_b, F_c \ll \kappa$, and the probability amplitude met the condition of $|C_{000}|\gg |C_{100}|,|C_{010}|,|C_{001}|\gg |C_{020}|$, so, we can set the $|C_{020}|=0$, $|C_{000}|=1$ Ref.~\cite{53}, and in order to calculate conveniently, we make the $\Delta_a=\Delta_c=\Delta$. According to the above conditions we can get a solution for Eqs.~(\ref{10})
 \begin{eqnarray}
&&(F_b^{2}+F_cg+F_aJ)+i(F_b^{2}\Delta+F_cg\Delta_b+F_aJ\Delta)=0,\nonumber\\
\label{11}
\end{eqnarray}
In this paper, we only consider the case $\Delta=\Delta_b=0$~\cite{01}, and we can get a conditional equation
\begin{eqnarray}
&&g=-\frac{F_b^{2}+F_a J}{F_c},\nonumber\\
\label{0071}
\end{eqnarray}
which is the optimal conditions for UPB for mode $b$. When the optimal conditions are satisfied, the photons cannot occupy the state $|020\rangle$, which can be examined by numerical simulation.

\begin{figure}[h]%"[]"中为位置参数，四个参数tbph 依次是置顶、置底 、浮动、当前位置，，选用的参数优先顺序为h-t-b-p
\centering
\includegraphics[scale=0.50]{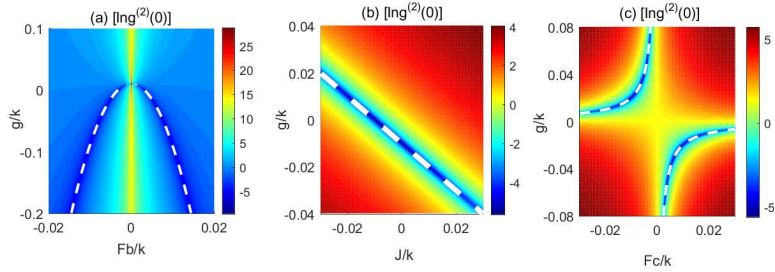}%"scale"后的数字为图形的宽度，也可用"width=1.0\columnwidth"定义
\caption{(Color online) (a) Logarithmic plot of the $g^{(2)}(0)$ as functions of $F_b/\kappa$ and $g/\kappa$, and we set $F_a/\kappa=0.01$, $F_c/\kappa=0.01$ and $J/\kappa=0.01$. (b) Logarithmic plot of the $g^{(2)}(0)$ varying with $J/\kappa$ and $g/\kappa$ for $F_a/\kappa=0.01$, $F_b/\kappa=0.01$ and $F_c/\kappa=0.01$. (c) Logarithmic plot of the $g^{(2)}(0)$ with $F_c/\kappa$ and $g/\kappa$, where $F_a/\kappa=0.01$, $F_b/\kappa=0.01$ ,$J/\kappa=0.01$. In both (a), (b) and (c) the dashed lines represent the analytical results.}
\label{fig2}
\end{figure}

In the following, we will study the UPB by the numerical simulation, and compare the results with the analytical solutions show in Eqs.~(\ref{0071}).
Under a truncated Fock space we can via solving the master equation numerically get the second-order correlation functions $g^{(2)}(0)$. In the current system, the Hilbert spaces are
truncated to five dimensions for cavity modes $a$ , $b$ and $c$, respectively.
For convenience, we rescale all parameters are in units of the dissipation rate $\kappa$ in Eqs.~(\ref{0071}), and the normalized optimal condition can be written as
\begin{eqnarray}
&&g/\kappa=-\frac{(F_b/\kappa)^{2}+F_aJ/\kappa^2}{F_c/\kappa},\nonumber\\
\label{0072}
\end{eqnarray}
Now, we numerically analyze the system to find the relationship between the driving strengths $Fa/\kappa$, $Fb/\kappa$, $Fc/\kappa$ and the Second order nonlinear coupling strength $g/\kappa$, $J/\kappa$ when the UPB occurs in cavity $b$. The results show in Fig.~\ref{fig2}.

In Fig.~\ref{fig2}(a), we show the logarithmic plot of the $g^{(2)}(0)$ varying with the $F_b/\kappa$ and $g/\kappa$, the other parameters are $F_a/\kappa=-0.01$, $F_c/\kappa=0.01$ and $J/\kappa=0.01$. We find that there is a parabolic dip corresponds to $g^{(2)}(0)\ll1$, and the white dotted line represented as the analytic condition from the equation of Eqs.~(\ref{0072}), which agree well with the numerical solution. According to existing parameters and the results show in Fig.~\ref{fig2}(a), when the strengths of the bimodal driving are the same the UPB can also happen, this is completely different from a system with a Kerr nonlinearity~\cite{01}, and when $F_a=F_b=F_c=F$, the optimal blocking condition will be simplified as
\begin{eqnarray}
&&g/\kappa=-(F/\kappa+J/\kappa),\nonumber\\
\label{0073}
\end{eqnarray}
In Fig.~\ref{fig2}(b) we logarithmic plot of the $g^{(2)}(0)$ varying with the $J/\kappa$ and $g/\kappa$ for $F_a/\kappa= F_b/\kappa=F_c/\kappa=0.01$. The blocking area appears as an oblique line, the white dotted line denote the analytic condition from the equation of Eqs.~(\ref{0072}), which is perfect fit with numerical solution. Here we set $F_a/\kappa= F_b/\kappa=F_c/\kappa$, so the analytic solution also satisfies equation of Eqs.~(\ref{0073}), consistent with the figure.
In Fig.~\ref{fig2}(c), we logarithmic plot of the $g^{(2)}(0)$ as the function of $g/\kappa$ and $F_c/\kappa$, where $F_a/\kappa=0.01$, $F_b/\kappa=0.01$ and $J/\kappa=0.01$. The white dotted lines denote the analytical condition, and since the $F_c$ appears in the denominator of the analytic condition, the analytic solution is hyperbolic, which is consistent with the figure, and the numerical solution is in perfect agreement with the analytical solution.

Next, the numerical and analytical solutions of the system will be further compared and analyzed, we will derive an analytic calculation expression for the $g^{(2)}(0)$, and compare it with the numerical simulations. In order to approximate get the analytic solution of $g^{(2)}(0)$, we solve Eqs.~(\ref{10}) under the weak driving condition and and the vacuum state $C_{000}$ approximately has unit occupancy, i.e., $C_{000}=1$. We can get a new coefficient equation as
\begin{eqnarray}
&&F_a +(\Delta_a-\frac{i\kappa}{2})C_{100}+\sqrt{2}J C_{020}=0,\nonumber\\
&&F_b +(\Delta_b-\frac{i\kappa}{2})C_{010}+\sqrt{2}F_b C_{020}=0,\nonumber\\
&&F_c +(\Delta_c-\frac{i\kappa}{2})C_{001}+\sqrt{2}gC_{020}=0,\nonumber\\
&&\sqrt{2}F_b C_{010}+\sqrt{2}J C_{100}+\sqrt{2}g C_{001}+2(\Delta_a-\frac{i\kappa}{2})C_{020}=0,
\label{0074}
\end{eqnarray}
For convenience, we set the $\Delta_a=\Delta_c=\Delta$, and the solutions of Eqs.~(\ref{0074}) are given as follows:
\begin{eqnarray}
&&C_{100}=\frac{8iJ(F_b^{2}+F_cgy)+2F_a(4g^{2}+xy)(-ik+2\Delta_b)}{xy(4g^{2}+4J^{2}+xy)},\nonumber\\
&&C_{010}=\frac{-2iF_b}{y},\nonumber\\
&&C_{001}=\frac{8iF_b^{2}gx+(-ik+2\Delta_b)[-4F_agJ+F_c(4J^{2}+xy)]}{xy(4y^{2}+4J^{2}+xy)},\nonumber\\
&&C_{020}=\frac{2\sqrt{2}[F_b^{2}x+y(F_cg+F_aJ)]}{xy(4g^{2}+4J^{2}+xy)},
\label{0075}
\end{eqnarray}
where the parameters $x=k+i2\Delta$ and $y=k+i2\Delta_b$ in the Eqs.~(\ref{0075}).
The $g^{(2)}(0)$ can be expressed as the probabilities of $n$-photon distribution function $P_n$ as
\begin{eqnarray}
&&g^{(2)}(0)=\frac{\sum_nn(n-1)P_n}{(\sum_nnP_n)^2},\nonumber\\
\label{0076}
\end{eqnarray}
and under the weak pumping limit the $g^{(2)}(0)$ can be written as
\begin{eqnarray}
&&g^{(2)}(0)\simeq\frac{2|C_{020}|^2}{|C_{010}|^4}.\nonumber\\
\label{0077}
\end{eqnarray}
According to Eqs.~(\ref{0075}) and Eqs.~(\ref{0077}) we can obtain the
\begin{eqnarray}
&&g^{(2)}(0)=\frac{(F_b^{2}+F_cg+FaJ)^{2}\kappa^{4}}{F_b^{4}(4g^{2}+4J^{2}+\kappa^{2})}
\label{0078}
\end{eqnarray}

\begin{figure}[h]%"[]"中为位置参数，四个参数tbph 依次是置顶、置底 、浮动、当前位置，，选用的参数优先顺序为h-t-b-p
\centering
\includegraphics[scale=0.60]{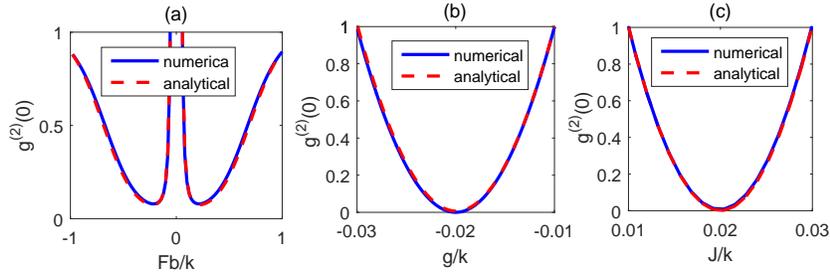}\nonumber\\%"scale"后的数字为图形的宽度，也可用"width=1.0\columnwidth"定义
\caption{(Color online) We plot of the $g^{(2)}(0)$ as functions of $F_b/\kappa$, $g/\kappa$ and $J/\kappa$. In fig(a) we set $F_a/\kappa=0.01$, $F_c/\kappa=0.1$, $J/\kappa=0.3$ and $g/\kappa=0.3$. In fig(b) $F_a/\kappa=0.01$ ,$F_b/\kappa=0.01$, $F_c/\kappa=-0.01$, and $g/\kappa=0.01$. In fig(c) $F_a/\kappa=0.01$ ,$F_b/\kappa=0.01$, $F_c/\kappa=0.01$, and $J/\kappa=0.01$. In both of (a), (b) and (c) the dashed lines indicate the analytical results.}
\label{fig3}
\end{figure}

In order to make a comparative analysis of $g^{(2)}(0)$ obtained by analytical calculation and numerical calculation, in Fig.~\ref{fig3}, we plot $g^{(2)}(0)$ versus various parameters, in which the red dotted line indicate the numerical solution and the blue solid line express the analytical solution.

In Fig.~\ref{fig3}(a), we plot the $g^{(2)}(0)$ versus with the $F_b/\kappa$, here $F_a/\kappa=0.01$, $F_c/\kappa=0.1$, $J/\kappa=0.3$, $g/\kappa=0.3$. The precise numerical solution in good agreement with the analytical solution, the minimum value of $g^{(2)}(0)$ appear at position of $F_b/\kappa=\pm0.16$, the results are agree well with the predicted based on the optimal blocking condition Eqs.~(\ref{0072}).
In Fig.~\ref{fig3}(b), we plot the $g^{(2)}(0)$ with the nonlinear interaction strength $g/\kappa$, where $F_a/\kappa=0.01$, $F_b/\kappa=0.01$, $F_c/\kappa=0.01$ and $J/\kappa=0.01$. The numerical solution agrees well with the analytical solution, and according to the optimal analytic condition Eqs.~(\ref{0072}) the optimum blocking position should appear at $g/\kappa=-0.02$, which in perfect agreement with the graph.
In Fig.~\ref{fig3}(c), we plot the $g^{(2)}(0)$ vs the nonlinear interaction strength $J/\kappa$, with $F_a/\kappa=0.01$, $F_b/\kappa=0.01$, $F_c/\kappa=-0.01$ and $g/\kappa=0.01$. The results are consistent with the above two figures, the numerical solution agrees well with the analytical solution, and according to the optimal analytic condition, the perfect blocking position happen at $J/\kappa=0.02$.
By comparing Fig.~\ref{fig3}(b) and Fig.~\ref{fig3}(c), we can find that the expression of the curves are same, it is just the difference in the location of the perfect block, which due to the different parameters set, and according to the Hamiltonian of the system Eqs.~(\ref{04}), these two nonlinear coupling terms are symmetric, the two nonlinear interaction strengths $g/k$ and $J/k$ play the same roles in this system, the results are in a good agreement with the two figures, meanwhile, the above conclusions are also agree well with the results shown in Fig.~\ref{fig2}(b).

\begin{figure}[h]%"[]"中为位置参数，四个参数tbph 依次是置顶、置底 、浮动、当前位置，，选用的参数优先顺序为h-t-b-p
\centering
\includegraphics[scale=0.50]{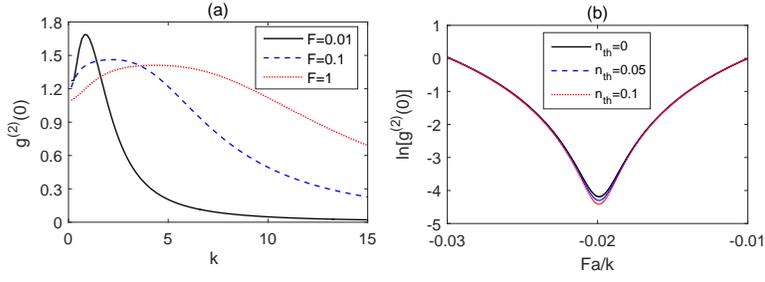}\nonumber\\%"scale"后的数字为图形的宽度，也可用"width=1.0\columnwidth"定义
\caption{(Color online) (a) Under the different driving strengths plot the $g^{(2)}(0)$ vs the dissipation rate $\kappa$, under the different driving strengths. Consider the optimal blocking condition in the black solid line $F_a/\kappa=F_b/\kappa=F_c/\kappa=0.01$, $J/\kappa=0.5$ and $g/\kappa=-0.51$, in the blue dotted line $F_a/\kappa=F_b/\kappa=F_c/\kappa=0.1$, $J/\kappa=0.5$ and $g/\kappa=-0.6$, and in the red point line $F_a/\kappa=F_b/\kappa=F_c/\kappa=1$, $J/\kappa=0.5$ and $g/\kappa=-1.5$. (b) Under the different number of thermal photons $\bar{n}_{th}$ logarithmic plot the $g^{(2)}(0)$ vs the driving strength $F_a/\kappa$, with $F_b/\kappa=0.01$, $F_c/\kappa=0.01$, $J/\kappa=0.01$ and $g/\kappa=0.01$ are same in the in these three lines, and in the black solid line we set $\bar{n}_{th}=0$, in the blue dotted line $\bar{n}_{th}=0.05$, in the red point line $\bar{n}_{th}=0.1$. Both of the (a) and (b) are numerical results.}
\label{fig4}
\end{figure}

Next, we discuss the effect of dissipation rate $\kappa$ and driving strength on UPB in the current system.
First, according to the
the optimal condition for strong antibunching Eqs.~(\ref{0071}), the $\kappa$ does not appear in the expression for the analytic condition, so, in theory, $\kappa$ has no effect on UPB. In Fig.~\ref{fig4} (a) we plot the $g^{(2)}(0)$ versus with the $\kappa$, in the black solid line we set $F_a/\kappa=F_b/\kappa=F_c/\kappa=0.01$, $J/\kappa=0.5$ and the value of $g/\kappa$ is calculated by the equation of Eqs.~(\ref{0071}) as $g/\kappa=-0.51$, strong UPB occurs in the black solid line, and the effect of UPB does not change significantly when $\kappa$ varies over a large range, and the same thing happens in the blue dotted line and red point line in Fig.~\ref{fig4}(a). So, the UPB effect is insensitive to $\kappa$ in this system, the results are consistent with that of theoretical prediction.
Secondly, in order to realize UPB theoretically, the condition of weak drive must be satisfied, we plot the $g^{(2)}(0)$ vs the dissipation rate $\kappa$, under the different driving strengths. For convenience, we set the driving strengths $F_a=F_b=F_c=F$ in the three lines, and in the blue dotted line the parameter are $F/\kappa=0.1$, $J/\kappa=0.5$ and according to optimal blocking conditions Eqs.~(\ref{0071}) the $g/\kappa=-0.6$, in the red point line $F/\kappa=1$, $J/\kappa=0.5$ and the $g/\kappa=-1.5$ is obtained by Eqs.~(\ref{0071}). By comparing these three curves we can find the UPB effect occurred in all three curves, but the blocking effect decreased significantly with the increase of $F/\kappa$, the result is agree well with theoretical prediction.
In the previous research, we study the UPB effect with zero temperature. Now, we will investigate the UPB effect with nonzero temperature. In order to discuss discuss the effect of temperature on the UPB, we need to redefine the density matrix $\hat{\rho}$ for this system as follows:
\begin{eqnarray}
\frac{\partial\hat{\rho}}{\partial t}&=&-i[\hat{H},\rho]+\frac{\kappa_a}{2}(\bar{n}_{th}+1)(2\hat{a}\hat{\rho}\hat{a}^\dag+\hat{a}^\dag\hat{a}\hat{\rho}+\hat{\rho}\hat{a}^\dag\hat{a})\nonumber\\
&&+\frac{\kappa_b}{2}(\bar{n}_{th}+1)(2\hat{b}\hat{\rho}\hat{b}^\dag+\hat{b}^\dag\hat{b}\hat{\rho}+\hat{\rho}\hat{b}^\dag\hat{b})\nonumber\\
&&+\frac{\kappa_c}{2}(\bar{n}_{th}+1)(2\hat{c}\hat{\rho}\hat{c}^\dag+\hat{c}^\dag\hat{c}\hat{\rho}+\hat{\rho}\hat{c}^\dag\hat{c})\nonumber\\
&&+\frac{\kappa_a}{2}\bar{n}_{th}(2\hat{a}^\dag\hat{\rho}\hat{a}+\hat{a}\hat{a}^\dag\hat{\rho}+\hat{\rho}\hat{a}\hat{a}^\dag)\nonumber\\
&&+\frac{\kappa_b}{2}\bar{n}_{th}(2\hat{b}^\dag\hat{\rho}\hat{b}+\hat{b}\hat{b}^\dag\hat{\rho}+\hat{\rho}\hat{b}\hat{b}^\dag)\nonumber\\ &&+\frac{\kappa_c}{2}\bar{n}_{th}(2\hat{c}^\dag\hat{\rho}\hat{c}+\hat{c}\hat{c}^\dag\hat{\rho}+\hat{\rho}\hat{c}\hat{c}^\dag),
\label{08}
\end{eqnarray}
In the Fig.~\ref{fig4}(b), we logarithmic plot the $g^{(2)}(0)$ as a function of $F_a/\kappa$, under the different number of thermal photons $\bar{n}_{th}$, where $F_b/\kappa=0.01$, $F_c/\kappa=0.01$, $J/\kappa=0.01$ and $g/\kappa=0.01$ are same in the in these three lines, and in the black solid line we set $\bar{n}_{th}=0$, in the blue dotted line $\bar{n}_{th}=0.05$, in the red point line $\bar{n}_{th}=0.1$, the results show that the strongest UPB point appears on $F_a/\kappa=1.3$, just as predicted. Moreover, when $\bar{n}_{th}$ changes in a large ranges, the PB does not change significantly, which indicate that this scheme is not sensitive to the change of the reservoir temperature, that make the system easier to implement experimentally.

\begin{figure}[h]%"[]"中为位置参数，四个参数tbph 依次是置顶、置底、浮动、当前位置，，选用的参数优先顺序为h-t-b-p
\centering
\includegraphics[scale=0.50]{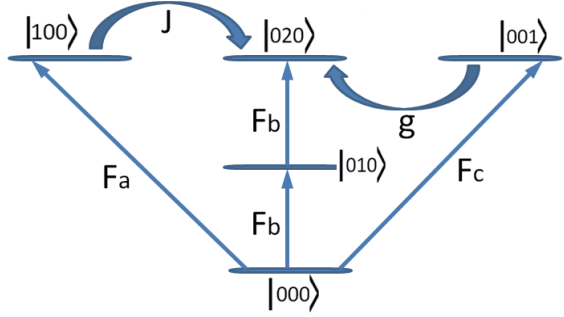}%"scale" 后的数字为图形的宽度，也可用"width=1.0\columnwidth"定义
\caption{Energy-level diagram.  The zero-, one-, and two-photon states (horizontal short lines) and the transition paths leading to the quantum interference responsible for the UPB (The arrows indicate the interference path).}
\label{fig5}
\end{figure}
Finally, we study the physical mechanism by which this system forms UPB, and the physics behind unconventional photon blockade is the effect of quantum interference between different paths. The energy-level structure and transition paths are shown in Fig.~\ref{fig5}. There are three paths for the system to reach the two-photon state of mode $b$:
(i) $|000\rangle\stackrel{\underrightarrow{F_b}}{}|010\rangle\stackrel{\underrightarrow{F_b}}{}|020\rangle$. (ii)$|000\rangle\stackrel{\underrightarrow{F_a}}{}|010\rangle\stackrel{\underrightarrow{J}}{}|020\rangle$.
(iii) $|000\rangle\stackrel{\underrightarrow{F_c}}{}|001\rangle\stackrel{\underrightarrow{g}}{}|020\rangle$.
When the optimal conditions for photon antibunching are satisfied, the photons come from different pathways destructive interference and the photons cannot occupy the state $|020\rangle$, the UPB will occur.

\section{Conclusions}
In this paper, we investigate the unconventional photon blockade in a three-mode system. The optimal condition for strong antibunching is found in the low frequency mode by analytical solutions and discussions of the optimal condition are presented. By comparing the numerical and analytical solutions, we find that they are in good agreement. By the analysis of numerical solutions, we can find this scheme is immune to the change of decay rates, and it is also insensitive to changes of reservoir temperature, in the experiments, the relatively large blocking windows are required, and the current three-mode scheme contains three driving terms and two second order nonlinear coupling terms make the system have more adjustable parameters to form a larger blocking window, both of which increases the possibility of experimental implementation. Therefore, the scheme can be used as a single photon source theoretically.
\section*{Acknowledgments}
This work is supported by the National Natural Science Foundation of China with Grants No.11647054. This work is also supported by the Science and Technology Development Program of Jilin province, China with Grant No.2018-0520165JH.
\section*{Disclosures:} The authors declare no conflicts of interest.\\

\bibliography{mybibfile}

\section*{References}
\begin{thebibliography}{99}


\bibitem{01} Y. H. Zhou, H. Z. Shen,  X. X. Yi. : Phys. Rev. A \textbf{92}, 023838 (2015).
\bibitem{02} A. J. Shields. :Nat. Photon \textbf{1}, 215 (2007).
\bibitem{03} L. Davidovich. :Rev. Mod. Phys \textbf{68}, 127 (1996).
\bibitem{04} V. Scarani, H. Bechmann-Pasquinucci and M. Peev. :Rev. Mod. Phys \textbf{81}, 1301 (2009).
\bibitem{05} V. Giovannetti, S. Lloyd, and L. Maccone. :Nat. Photon \textbf{5}, 222 (2011).
\bibitem{06} E.Knill, R.Lafiamme, and G.J.Milburn. :Nature(London) \textbf{409}, 46 (2001).
\bibitem{10} X. Gu, A. F. Kockum, A. Miranowicz, Y.X. Liu, and F. Nori. :Phys. Rep \textbf{1}, 718-719 (2017).
\bibitem{11} I. Buluta, S. Ashhab, and F. Nori. :Rep. Prog. Phys \textbf{74}, 104401 (2011).
\bibitem{12} C. Lang, D. Bozyigit, C. Eichler. :Phys. Rev. Lett \textbf{106}, 243601 (2011).
\bibitem{13} A. J. Hofiman, S. J. Srinivasan, S. Schmidt, L. Spietz, J. Aumentado and A. A. Houck. :Phys. Rev Lett \textbf{107}, 053602 (2011).
\bibitem{14} S. Ferretti, L. C. Andreani, and D. Gerace. :Phys. Rev. A \textbf{82}, 013841 (2010).
\bibitem{15} J. Q. Liao and C. K. Law. :Phys. Rev. A \textbf{82}, 053836 (2010).
\bibitem{16} A. Miranowicz, M. Paprzycka, Y.X. Liu, J. Bajer, and F. Nori. :Phys. Rev. A \textbf{87}, 023809 (2013).
\bibitem{17} H. Xie, G. W. Lin, X. Chen, Z. H. Chen, and X. M. Lin. :Phys. Rev. A \textbf{93}, 063860 (2016).
\bibitem{18} C. Zhai, R. Huang, B. Li, H. Jing, and L.M. Kuang. :arXiv \textbf{1901}, 07654 (2019).
\bibitem{19} A. Imamoglu, H. Schmidt, G. Woods, and M. Deutsch. :Phys. Rev. Lett \textbf{79}, 1467 (1997).
\bibitem{z} Y. H. Zhou, H. Z. Shen, X. Y. Zhang, and X. X. Yi. :Phys. Rev. A \textbf{97}, 043819 (2018).
\bibitem{ab} T. C. H. Liew and V. Savona. :Phys. Rev. Lett \textbf{104}, 183601 (2010).
\bibitem{20} R. J. Brecha, P. R. Rice, and M. Xiao. :Phys. Rev. A \textbf{59}, 2392 (1999).
\bibitem{22} P. Rabl. :Phys. Rev. Lett \textbf{07}, 063601 (2011).
\bibitem{24} J. Q. Liao and F. Nori. :Phys. Rev. A \textbf{88}, 023853 (2013).
\bibitem{25} S. D. Bennett, K. Stannigel, S. J. M. Habraken, P. Rabl, P. Zoller, and M. D. Lukin. :Phys. Rev. A \textbf{87}\, 013839 (2013).
\bibitem{26} A. Majumdar and D. Gerace. :Phys. Rev. B \textbf{87}, 235319 (2013).
\bibitem{27} H. Z. Shen, Y. H. Zhou, and X. X. Yi. :Phys. Rev. A \textbf{90}, 023849 (2014).
\bibitem{e} H. Z. Shen, Y. H. Zhou, X. X. Yi. :Phys. Rev. A \textbf{91}, 063808 (2015).
\bibitem{28} D. E. Chang, A. S. Sorensen, E. A. Demler, and M. D. Lukin. :Nat. Phys \textbf{3}, 807 (2007).
\bibitem{29} D. Gerace, H. E. T. ureci, V. Giovannetti, and R. Fazio. :Nat. Phys \textbf{5}, 281 (2009).
\bibitem{30} F. Fratini, E. Mascarenhas, L. Safari, J-Ph. Poizat, D. Valente, D. Gerace, and M. F. Santos. :Phys. Rev. Lett \textbf{113}, 243601 (2014).
\bibitem{31} E. Mascarenhas, D. Gerace, D. Valente, S. Montangero,and M. F. Santos. :Europhys. Lett \textbf{106}, 54003 (2014).
\bibitem{33} M. Bayer, T. Gutbrod, J. P. Reithmaier, A. Forchel, T. L. Reinecke, P. A. Knipp, A. A. Dremin, and V. D. Kulakovskii. :Phys. Rev. Lett \textbf{81}, 2582 (1998).
\bibitem{34} Y. P. Rakovich and J. F. Donegan. :Laser Photon. Rev \textbf{4}, 179 (2010).
\bibitem{c} H. Z. Shen, S. Xu, Y. H.Zhou, G. Wang, X. X. Yi. :J. Phys. B \textbf{51}, 035503 (2018).
\bibitem{b} H. Z. Shen, Y. H. Zhou, H. D. Liu, G. C. Wang, and X. X. Yi. :Opt. Express \textbf{23}, 32835 (2015).
\bibitem{d} H. Z.Shen, C. Shang, Y. H. Zhou, X. X. Yi. :Phys. Rev. A \textbf{98}, 023856 (2018).
\bibitem{f} Y. H. Zhou, H. Z. Shen,  X. Q. Shao, X. X. Yi. :Opt. Express \textbf{24}, 17332 (2016).
\bibitem{35} M. Bamba, A. Imamofiglu, I. Carusotto, and C. Ciuti. :Phys. Rev. A \textbf{83}, 021802(R) (2011).
\bibitem{36} I. Carusotto and C. Ciuti. :Rev. Mod. Phys \textbf{85}, 299 (2013).
\bibitem{37} A. Majumdar, M. Bajcsy, A. Rundquist. :Phys. Rev. Lett \textbf{108}, 183601 (2012).
\bibitem{38} W. Zhang, Z. Y. Yu, Y. M. Liu, and Y. W. Peng. :Phys. Rev. A \textbf{89}, 043832 (2014).
\bibitem{39} M. Bamba and C. Ciuti. :Appl. Phys. Lett \textbf{99}, 171111 (2011).
\bibitem{40} X. W. Xu and Y. J. Li, J. Opt . B :At. Mol. Opt. Phys \textbf{46}, 035502 (2013).
\bibitem{42} S. Ferretti, V. Savona, and D. Gerace. :New J. Phys \textbf{15}, 025012 (2013).
\bibitem{43} H. Flayac and V. Savona. :Phys. Rev. A \textbf{88}, 033836 (2013).
\bibitem{44} D. Gerace and V. Savona. :Phys. Rev. A \textbf{89}, 031803(R) (2014).
\bibitem{45} H. Z. Shen, Y. H. Zhou, and X. X. Yi. :Phys. Rev. A \textbf{90} 023849 (2014).
\bibitem{46} D. Sanvitto and S. Kena-Cohen, Nat. Mater \textbf{15}, 1061 (2016).
\bibitem{47} H. Flayac and V. Savona. :Phys. Rev. A \textbf{95}, 043838 (2017).
\bibitem{50} Y. H. Zhou, Q. C. Wu, B. L. Ye, L. Y. Xue and H. Z. Shen. :Int. Jou. Theor. Phys \textbf{58}, 427-479 (2019).
\bibitem{i}S. L. Su, Y. Z. Tian,  H. Z. Shen, H. P. Zang, E. J. Liang, S. Zhang. :Phys. Rev. A \textbf{96}, 042335 (2017).
\bibitem{j} S. L.Su, Y. Gao, E. J. Liang, S. Zhang. :Phys. Rev. A \textbf{95}, 022319 (2017).
\bibitem{n}S. L. Su, E. J Liang, S. Zhang, J. J. Wen, L. L. Sun, Z. Jin , A.D. Zhu. :Phys. Rev. A \textbf{93}, 012306 (2016).
\bibitem{53} Y. X. Liu, X. W. Xu, A. Miranowicz, and F. Nori. :Phys. Rev. A \textbf{89}, 043818 (2014).
\end{thebibliography}

\end{document}